\documentclass[%
 reprint,prl
]{revtex4-2}

\usepackage{graphicx}
\usepackage{dcolumn}
\usepackage{bm}
\usepackage{hyperref}
\usepackage{amsmath}
\usepackage{amssymb}
\usepackage{amsfonts}
\usepackage{tikz}
\usepackage{tikz-cd}
\usepackage{colortbl}
\usepackage{array}
\usepackage{mymacros}  
\usepackage{bbm} 
\usepackage{xcolor} 
\hypersetup{colorlinks=true, linktocpage=true}

\definecolor{green}{rgb}{0.7, 0.9, 0.7}
\definecolor{blue}{rgb}{0.7, 0.8, 0.9}
\definecolor{yellow}{rgb}{1.0, 1.0, 0.8}
\definecolor{red}{rgb}{1.0, 0.7, 0.7}

\definecolor{softgreen}{rgb}{0.7, 0.9, 0.7}
\definecolor{softblue}{rgb}{0.7, 0.8, 0.9}
\definecolor{softyellow}{rgb}{1.0, 1.0, 0.8}
\definecolor{softred}{rgb}{1.0, 0.7, 0.7}
\newcommand*\circled[2]{%
  \tikz[baseline=(char.base)]{
    \node[shape=circle,fill=#1,draw,inner sep=2pt, minimum size=12pt] (char) {#2};}}

\begin{document}


\title{On topological defects in Chern-Simons theory}

\author{Alex S. Arvanitakis$^{1}$, Lewis T. Cole$^{2}$, Saskia Demulder$^{3}$, and Daniel C. Thompson$^{2}$  \\
\noindent \vspace{-5pt} \\
        \small{\textit{$^{1}$Division of Theoretical Physics, Rudjer Bo\v{s}kovi\'c Institute,
Bijeni\v{c}ka 54, 10000 Zagreb, Croatia}}\\ 
    \small{\textit{$^{2}$Department of Physics, Swansea University, Swansea SA2 8PP, United Kingdom}} \\
        \small{\textit{ $^{3}$Department of Physics, Ben-Gurion University of the Negev}},\\ \small{\textit{ David Ben Gurion Blvd 1, Beer Sheva 84105, Israel}} \\
}





\date{\today}

\begin{abstract}
We construct a new class of topological surface defects in Chern-Simons theory with non-compact, non-Abelian gauge groups. These defects are characterized by isotropic subalgebras defined by solutions of the modified classical Yang-Baxter equation, and their fusion realizes a semi-group structure with non-invertible elements.  From a Hamiltonian perspective, we calculate this fusion using the composition of Lagrangian correspondences within the Weinstein symplectic category. Applications include   boundary terms and conditions in $AdS_3$ gravity and higher-spin theories.
\end{abstract}

\maketitle


\section{\label{sec:Introduction} Introduction}
  A topological defect may separate  one phase of a physical system from its symmetry-transformed counterpart.   The topological nature of the defects enables fusion which captures the composition law of      \emph{generalized symmetries} \cite{Gaiotto:2014kfa}  (see reviews \cite{Schafer-Nameki:2023jdn, Luo:2023ive}). For conventional symmetries, this corresponds to group multiplication. More generally, fusion results in a superposition of defects   yielding a non-invertible composition law.  
 
In this work, we exhibit topological defects in 3D Chern-Simons (CS) theory with a novel and perhaps more elementary kind of non-invertibility: where fusion closes on single elements but no inverse exists, resulting in a \emph{semi-group} composition law. These defects are obtained via the folding trick from topological boundary conditions associated with \emph{lagrangian} (maximal isotropic) subalgebras of   a ``doubled''  Lie algebra $\mathfrak{d} = \mathfrak{g} \oplus \mathfrak{g}$.

For abelian CS theory where $\mathfrak{g} = \mathfrak{u}(1)^d$ our characterisation of topological defects agrees with that of Kapustin and Saulina \cite{Kapustin:2010hk}. When the inner product $\kappa$ is positive-definite, we make contact with certain defect actions studied by Roumpedakis et al.~\cite{Roumpedakis:2022aik} via the introduction of edge modes; also, we show fusion corresponds to the orthogonal group  $O(\kappa)$.
However, we show that the more general case of indefinite  $\kappa$ admits non-invertible defects forming a semi-group.  This is illustrated with the examples $\kappa = \mathrm{diag}(+1, -1)$ and $\kappa = \mathrm{diag}(+1, +1, -1)$.

We then construct defects for non-abelian
Chern-Simons theory, addressing a longstanding open problem identified in \cite{Kapustin:2010hk} (see also \cite{Kapustin:2010if,Fliss:2020cos}).  Given non-Abelian $\fg$, we construct lagrangians from $\mathcal{R}$-matrices solving the modified Yang-Baxter equation. These $\mathcal{R}$-defects, specific to non-compact $\mathfrak{g}$, are shown to form a semi-group under fusion using the  Hamiltonian approach of \cite{Arvanitakis:2023sho}. Such constructions have significant applications: CS theory on $\mathfrak{g}^\mathbb{C}$ has been extensively studied \cite{Witten:1989ip}, while theories with $\mathfrak{g}$ as a Drinfel’d double provide a TQFT framework for T-duality and Poisson-Lie T-duality \cite{Severa:2016prq,Arvanitakis:2021lwo}. 

The Chern-Simons action with $\mathfrak{g} = \mathfrak{sl}_2 \oplus \mathfrak{sl}_2$ provides a rewriting of the 3D Einstein-Hilbert action with negative cosmological constant  \cite{Achucarro:1986uwr,Witten:1988hc}.  Augmenting the action with the boundary term for the Drinfel'd-Jimbo $\cR$-matrix reproduces the full gravitational boundary contributions used in \cite{Apolo:2015zxh} and readily generalizes to higher-spin theories. 
We also introduce novel boundary conditions associated to the $\cR$-matrix and show these include asymptotically AdS spacetimes  whose asymptotic symmetry algebra is identified as a single  Virasoro.

\section{\label{sec:Folding} Folding and Defects }

Chern-Simons theory is defined by the action  
\begin{equation}
    S_{\text{CS}}[A] =   \int_{M} \bigg( \biprod{A}{\dr A} + \frac{1}{3} \triprod{A}{A}{A} \bigg) \ ,
\end{equation}
where $\kappa_\fg= \biprod{\bcdot}{\bcdot}$ is an appropriately quantized  bilinear form on the algebra $\fg$.  We bisect \(M\) into northern, \(M_N\), and southern, \(M_S\), regions, with a shared boundary \(D\) of opposite orientations. Our goal is to define a theory that mediates interactions between  gauge fields \(A_{N,S}\) living in $M_{N,S}$ respectively  and any degrees of freedom localized on \(D\). If the full system is invariant under displacements of \(D\), the defect is said to be \emph{topological}.

We use the folding trick: since \( M_N \) and \( M_S \) are diffeomorphic near \( D \) we fold them with a parity-odd map \(\varphi: M_N \to M_S\).  
This reduces the problem  to an analysis of topological boundary conditions in the  \emph{doubled} CS action $ S_{\text{CS}}[\mathbb{A} ]  $ on $M_N$ with boundary $\pd M_N= D$ for a connection \(\bA = (A_N, A_S)\) valued in $  \fd = \fg_{N} \oplus  \fg_{S}$ 
with the bilinear form $\kappa_\fd= \biprodb{\bullet }{\bullet} =\kappa_{\fg_N} -  \kappa_{\fg_S}$. 

The boundary condition should ensure the vanishing of the surface term \( \int_D \biprodb{\delta \mathbb{A}}{\mathbb{A}} \) arising from the variation of the action.
 A common choice is \( \bA \vert_D = \star (\bA \vert_D) \), leading to chiral dynamics on \( D \), but this introduces a Hodge structure, which is not topological. Instead we require that the gauge field be valued in a \emph{lagrangian} subalgebra \( \fh \subset \mathfrak{d} \) \cite{Kapustin:2010hk} so that    $\bA \vert_{D} \in \Omega^{1} (D) \otimes \fh$.

Let \( P \) be a projector with complement \( P^\perp = \id - P \), with \( \textrm{im}(P) = \mathfrak{h} \) such that   \( P^\perp \bA|_D = 0 \). We augment the theory with a boundary term
\begin{equation}
\label{eq:Stot}
    S_{\text{tot}}[\mathbb{A}] = S_{\text{CS}}[\mathbb{A}] +  \int_D \biprodb{\bA}{P^\perp \bA} \, .
\end{equation}
 This introduces a boundary interaction between the north and south theories while respecting the boundary condition, as the surface term in \( \delta S_{\text{tot}} \) is \( 2 \int_D \biprodb{\delta \mathbb{A}}{P^\perp \mathbb{A}} \). The boundary term is included to interpret \( P^\perp \bA|_D = 0 \) as a ``boundary equation of motion".

After including this boundary term half of the gauge invariance, \( \mathbb{A} \mapsto \mathbb{A}^\mathbbm{g} = \mathbbm{g}^{-1} \dr \mathbbm{g} + \mathbbm{g}^{-1} \mathbb{A} \mathbbm{g} \), is preserved.  The broken symmetry may be repaired by introducing St\"{u}ckelberg fields \( \mathbbm{h} = (h_N, h_S) \) transforming as \( \mathbbm{h} \mapsto \mathbbm{g}^{-1} \mathbbm{h} \). Then \( \mathbb{A}^{\mathbbm{h}} \) is invariant and \( \mathbbm{h} \) only appears in defect-localized terms outside of a Wess-Zumino term:
\begin{equation}\label{eq:fullmonkey} \begin{split}
         S_{tot}[\mathbb{A}^{\mathbbm{h}}] = S_{tot}[\mathbb{A}]+S_{WZ}[\mathbbm h]+ \int_D \biprodb{\mathbb A}{\dr\mathbbm h \mathbbm h^{-1}}\\+ \int_D  \biprodb{\mathbb A + \dr \mathbbm h\mathbbm h^{-1}}{(\operatorname{Ad}_{\mathbbm h} P^\perp \operatorname{Ad}_{\mathbbm h}^{-1}) (\mathbb A + \dr \mathbbm h\mathbbm h^{-1})} \, ,
\end{split}
\end{equation}
where we use the adjoint action $\operatorname{Ad}_{\mathbbm h}\cdot= \mathbbm h\cdot \mathbbm h^{-1}$. This action generalizes the Lagrangian description of condensation surfaces in abelian CS theory \cite[sec.~6.2.2]{Roumpedakis:2022aik}. 

We address fusion by dividing \( M \) into three regions: north \( N \), south \( S \), and equatorial belt \( I \). The folded fields are \( \mathbb{A}_{NI} = (A_N, A_I) \) and \( \mathbb{A}_{IS} = (A_I, A_S) \), with defects on \( D_{NI} = M_N \cap M_I \) and \( D_{IS} = M_I \cap M_S \) prescribed by projectors \( P_{NI} \) and \( P_{IS} \) into Lagrangians \( \mathfrak{h}_{NI} \) and \( \mathfrak{h}_{IS} \). Fusion is achieved by shrinking the equatorial belt, merging the defects \( D_{NI} = D_{IS} = D \), such that \( A_I \)  survives only as a non-dynamical field on $D$.

Terms involving \( A_I \) are of the form
\begin{equation}\label{eq:SI}
   S_{I}=  \int_D \frac{1}{2}  \biprod{A_I}{ Y A_I } +  \biprod{A_I }{ B}\, , 
\end{equation}
with \( Y \)  a constant skew matrix dependent on the input projectors, and \( B \) a one-form depending on the remaining fields. If \( Y \) is invertible \footnote{If \( \dim \mathfrak{g} \) is odd, \( Y \) has a null space. Whilst components of \( A_I \) transverse to this can be eliminated, the remainder \( A_I \) enforces the projection into the null space as a Lagrange multiplier. This phenomenon also occurs in even dimensions for some \( P_{NI} \) and \( P_{IS} \). Despite differences in the resulting defect actions, the on-shell boundary variations after fusion requires \( \mathbb{A}_{NS} \) to take values in \( \mathfrak{h}_{NS} \).}, we can eliminate \( A_I \), yielding
\begin{equation}
   S_{I}=  \int_D   - \frac{1}{2} \biprod{B }{ Y^{-1} B}  \, . 
\end{equation}
We will give an explicit example for $Y$ and $B$ in the following sections. 
The fused theory takes the form of \( S_{\text{tot}}[\mathbb{A}_{NS}] \) of eq.~\eqref{eq:Stot}, with a new projector \( P_{NS} \), imposing that \( \mathbb{A}_{NS} \) takes values in the lagrangian \( \mathfrak{h}_{NS} \).

Within this treatment,  calculating the resulting $\mathfrak{h}_{NS}$  from $\mathfrak{h}_{NI}$ and  $\mathfrak{h}_{IS}$ is  somewhat opaque.  To resolve this we now turn to a Hamiltonian  approach.   

\section{ A Hamiltonian Approach to Fusion}
Recent work \cite{Arvanitakis:2023sho} by one of us demonstrated that there is a 1-to-1 relation between lagrangian correspondences  and topological defects in Hamiltonian mechanics. We explain how this applies to CS theory and determine a composition rule for defect fusion.

The dynamics of a particle defines a curve   $\gamma(\tau)$ on phase space   \( {\cal M} = T^*Q \), with action $\int \;\gamma^\star \theta$ 
involving the tautological symplectic potential \( \theta = p \cdot dq \). A canonical transformation defines new coordinates \( (\tilde{p}, \tilde{q}) \) which we implement  via a worldline defect joining two paths \( \gamma \) and \( \tilde{\gamma} \) at fixed time $\tau'$. The folding trick in this context yields a path  $\Gamma(\tau) = (\gamma(\tau- \tau^\prime) ,\tilde{\gamma}(\tau^\prime - \tau) )$ in the space ${\cal M} \times \widetilde{{\cal M}}$ with ``doubled'' action $\int \Gamma^\star(\theta - \tilde{\theta})$ together with an appropriate boundary condition.  The defect is topological, i.e.~independent of \( \tau' \),  because the map $\mathcal M\to \tilde{\mathcal M}$ is a symplectomorphism. 

This is an example of a {\em lagrangian correspondence} in the sense of Weinstein's symplectic `category' \cite{weinstein1982symplectic}: the graph of the map between the symplectic manifolds \( ( {\cal M} , \omega) \) and \( (\widetilde{ {\cal M} }, \tilde{\omega}) \) defines a lagrangian submanifold \( L \) in the \emph{correspondence space} \( ( {\cal M}  \times \widetilde{ {\cal M} }, \omega - \tilde{\omega}) \), on which $\omega - \tilde{\omega}$ vanishes.  Correspondences  \(L_{ij} \subset {\cal M}_i \times {\cal M}_j\)  admit compositions   
\begin{equation}\label{eq:lagcorfusion} 
L_{ij} \circ L_{jk} = \Pi_{ik} \left[ (L_{ij} \times L_{jk}) \cap  ({\cal M}_i \times \Delta_j \times  {\cal M}_k) \right],
\end{equation}
where \( \Pi_{ik} \) projects onto \(  {\cal M}_i \times  {\cal M}_k \), and \( \Delta_j \) is the diagonal embedding \(  {\cal M}_j \hookrightarrow  {\cal M}_j \times  {\cal M}_j \).   This provides the fusion rules for topological defects in this context \cite{Arvanitakis:2023sho}. 

These arguments apply to the folded Chern-Simons theory in the neighbourhood of a defect $D$ where the spacetime $M$  takes the form $ \mathbb{R}_\tau    \times D$ without loss of generality.  After splitting $\mathbb{A} =   \mathbb{A}_0 \dr \tau +   \alpha$ with respect to `time' $\tau$ normal to the defect, \( \mathbb{A}_0 \) serves as a Lagrange multiplier enforcing Gauss's law, and the remaining action is that of a particle moving on the symplectic manifold  ${\cal M}(\frak{d}, D)$ of   $\frak{d}$-valued connections:
\begin{equation}
\begin{split}
    S_{CS}[\mathbb{A} ] = \int \bigg(   \Gamma^* \theta +    \int_D  \biprodb{ \mathbb{A}_0 \dr \tau  }{   F[\alpha] }  \bigg) \, , \\ 
   \theta = \int_D    \biprodb{ \alpha }{  \delta \alpha  } \, , \quad  \omega =\delta \theta =  \int_D   \biprodb{\delta  \alpha }{  \delta \alpha  } \, . 
\end{split}
\end{equation}
As  $\frak{d} = \frak{g}_N \oplus \frak{g}_S$ is equipped with the split signature pairing we view ${\cal M}(\frak{d}, D) $ as the correspondence space $  {\cal M}(\frak{g}_N, D) \times {\cal M}(\frak{g}_S, D) $.

Ignoring the Gauss law constraint initially, defects are identified with lagrangian submanifolds in \( {\cal M}(\mathfrak{d}, D) \). Such defects are topological with respect to displacement in the \( \tau \)-direction chosen above. Defects invariant under arbitrary infinitesimal diffeomorphisms correspond to lagrangians in \( {\cal M}(\mathfrak{d}, D) \) of the specific form \( \Omega^1(D) \otimes \mathfrak{h} \), where \( \mathfrak{h} \subset \mathfrak{d} \) is a maximal isotropic subspace.  Compatibility with the Gauss law implies \( \mathfrak{h} \) is closed, i.e., a lagrangian subalgebra \footnote{What must be true is that $\dr \bA +\bA^2=0$  on $D$ where $\bA$ takes values in $\frak h$. This is because symplectic reduction onto the moduli space of flat connections is a lagrangian correspondence; see \cite{Arvanitakis:2023sho} for details.} . Thus \( \mathbb{A}_{NS} \) restricts, on $D$, to a gauge field valued in \(  \mathfrak{h} \), up to gauge transformations.

The fusion of lagrangian correspondences given by eq.~\eqref{eq:lagcorfusion} simplifies to a composition of Lie algebras:
\begin{equation}\label{eq:liefusion}
\mathfrak{h}_{NI} \circ \mathfrak{h}_{IS} = \Pi_{NS} \left[ (\mathfrak{h}_{NI} \times \mathfrak{h}_{IS}) \cap \mathfrak{g}_N \times \Delta_{\mathfrak{g}_I} \times \mathfrak{g}_S \right]  \, , 
\end{equation}   
where \( \Delta_{\mathfrak{g}_I} \) is the diagonal embedding \( \mathfrak{g}_I \hookrightarrow \mathfrak{g}_I \times \mathfrak{g}_I \). Algorithmically, if \( v_{NI} = (v_N, v_I) \in \mathfrak{h}_{NI} \) and \( u_{IS} = (u_I, u_S) \in \mathfrak{h}_{IS} \),  eq.~\eqref{eq:liefusion} tells us to consider elements of the form \( (v_N, u_S) \) subject to the relation \( v_I = u_I \).  This operation is the associative, but non-abelian,  \emph{composition of relations} in set theory.

\section{\label{sec:Abelian} Abelian Defects}
Let  \( \mathfrak{g}_N =\fg_S= \mathfrak{u}(1)^d \) with the definite inner product \( \kappa_N = \kappa_S = \textrm{diag}(+,+,\dots,+) \). A family of lagrangians parameterised by  a skew-symmetric matrix \( \beta \) is 
\begin{equation}\label{eq:Deltabeta} \mathfrak{h}_\beta =  \{(Q_+ x, Q_- x) \in \fd \mid x\in\fg\}, \quad Q_\pm = \id \pm \beta \kappa  .
\end{equation} 
This contains the diagonal lagrangian \( \mathfrak{h}_{0} =  
\{(x, x)\in \fd \mid x\in \fg \}  \).   
A projector \footnote{In this presentation we set the kernels of the projectors $P_{NI}$ and $P_{IS}$ to be the anti-diagonal,   though the result applies more generally to any bi-vector transformation of the kernel. Both the image and kernel of \( P_{NS} \) are generated by fusion of the respective lagrangians eq.~\eqref{eq:liefusion}.} whose image is \( \mathfrak{h}_\beta \) is given by 
\begin{equation} 
P_\beta = \frac{1}{2} \begin{pmatrix}  Q_+ & Q_+ \\ Q_- & Q_- 
\end{pmatrix}\, .
\end{equation}
As \( Q_\pm \) are invertible, we can use the Cayley transform \( M_\beta = Q_+ Q_-^{-1} \in SO(d) \) to  write \( \mathfrak{h}_\beta =   \{ (M_\beta x,x) \} \). In fact, the space of lagrangians is $O(d,\mathbb R)$.  

Consider fusing defects with \( P_{NI} = P_\beta \) and \( P_{IS} = P_{\tilde{\beta}} \). In the Hamiltonian approach, applying eq.~\eqref{eq:liefusion} yields
\begin{equation}\label{eq:abelianfusion} 
\mathfrak{h}_{\beta} \circ \mathfrak{h}_{\tilde{\beta}} =  \{ (M_\beta \cdot M_{\tilde{\beta}} x, x) \mid x\in \fg \} \, ,
\end{equation}
i.e., fusion realises the group multiplication law in \( O(d) \). In the Lagrangian approach to fusion, the data defining the intermediate action in eq.~\eqref{eq:SI} are
\begin{equation}
Y = \beta + \tilde{\beta}, \quad B = Q_- A_N - \widetilde{Q}_+ A_S \, .
\end{equation}
Performing the elimination of $A_I$ we   recover a defect action defined by a projector \( P_{NS} \) whose image matches eq.~\eqref{eq:abelianfusion}, and whose kernel is the diagonal \( \mathfrak{h}_0 \).

A very different feature becomes apparent  when   $\kappa$ is \emph{indefinite}. Whilst $\mathfrak{h}_\beta$ of eq.~\eqref{eq:Deltabeta} remains a lagrangian,   $Q_\pm $ need not be invertible. Not coincidentally, such $\frak h_\beta$ can never fuse into the diagonal (identity) defect.

{\bf Example: $d=2$ with $\kappa=\textrm{diag} (+,-)$.}  In this setting   eq.\eqref{eq:Deltabeta}  defines   a one-parameter   family of lagrangians  
\begin{equation}
\mathfrak{h}_\beta    =  \{ (u + \beta v, v +\beta u, u- \beta v , v  -\beta u  ) \mid u,v\in\mathbb R \} \, .
\end{equation}
Here, and in the sequel, we denote elements of a lagrangian by their coordinates in the basis where $\kappa_N-\kappa_S=\textrm{diag}(+--+)$; $\fg_N$ being spanned by the first two.

For $\beta \neq \pm 1$  we use an $O(1,1)$ Cayley transform to rewrite $\mathfrak{h}_\omega     =  \{ ( \cosh \omega u + \sinh \omega  v,   \sinh \omega u + \cosh \omega v ,  u , v ) \} \, ,$ for which fusion acts by addition on the  rapidities  defined by $\beta = \tanh \omega / 2 $.  

However at $\beta = \pm 1 $ (i.e. in the infinite rapidity limit) we  find four lagrangians of the form 
\begin{equation}
\label{eq:2dlagrangians}
    {\mathfrak{a}}^\pm =  \{  (u, \pm u , v , \mp v )   \}  \, , \;\; {\mathfrak{b}}^\pm =  \{  (u, \pm u , v , \pm v )  \} \, . 
\end{equation} 
These lagrangians form a direct product of two two-element rectangular band  semi-groups, $  \mathbb{B}_2 = \{\mathfrak a^\pm,\mathfrak b^\mp\}$,  with fusion given by  (to be read as $\mathfrak a^-\circ\mathfrak a^+=\mathfrak b^-$): 
\begin{equation}
    \begin{array}{c|c c c c} 
    \circ  & {\mathfrak{a}}^+ & {\mathfrak{a}}^- &  {\mathfrak{b}}^+ &   {\mathfrak{b}}^- \\
    \hline 
     {\mathfrak{a}}^+ &  {\mathfrak{a}}^+ &  {\mathfrak{b}}^+    &{\mathfrak{b}}^+  & {\mathfrak{a}}^+  \\ 
       {\mathfrak{a}}^- &   {\mathfrak{b}}^-  &    {\mathfrak{a}}^- & {\mathfrak{a}}^-  &  {\mathfrak{b}}^- \\ 
            {\mathfrak{b}}^+ & {\mathfrak{a}}^+ &  {\mathfrak{b}}^+  &{\mathfrak{b}}^+  &  {\mathfrak{a}}^+  \\ 
                 {\mathfrak{b}}^- &  {\mathfrak{b}}^- &  {\mathfrak{a}}^-   &{\mathfrak{a}}^-  &   {\mathfrak{b}}^- \\  
    \end{array}  
\end{equation}   
One can never obtain the diagonal/identity defect $\mathfrak{h}_{0}$ from fusions of $  {\mathfrak{a}}^\pm$ or ${\mathfrak{b}}^\pm  $ with each other or indeed with $\fh_\omega$; they are non-invertible.

{\bf Example: $d=3$ with $\kappa=\textrm{diag}(+,+,-)$.}
 Away from the locus $1=\beta_1^2 +\beta_2^2 - \beta_3^2$, where  $\beta^{ab} = \epsilon^{abc}\beta_c$,  we can take $M_{\beta} =Q_+ Q_-^{-1}  $ as an $SO(2,1)$ matrix and construct invertible lagrangians as before. On   that locus we   have ($\beta$-dependent) elements   $t_\pm \in \fg $  obeying
\begin{equation}
 Q_\pm t_\pm = 0 \, , \quad Q_\pm t_\mp = 2 t_\mp \, , \quad \kappa (t_\pm,t_\pm) = 0 \, . 
\end{equation}
We can express a generic element $x   =  u t^+ +  
 v t^- + w t^\perp $ where      $t^\perp$ is $\kappa$-orthogonal to $t^\pm$.  This gives rise to non-invertible lagrangians of the  form  
 \begin{equation} 
    \frak h_\beta    =   \textrm{span} \{ \left( u t^+ +w t^\perp   ,   v t^- + w  t^\perp  \right) \} \, .   
\end{equation}
Fusion of these results in a slightly more general type of lagrangian (not necessarily of the form   $\frak h_\beta$): 
 \begin{equation} 
    \frak h_\beta  \circ   \frak h_{\tilde{\beta}}  =   \textrm{span} \{ \left( u t^+ +w t^\perp   ,   \tilde{v} \tilde{t}^- + w  \tilde{t}^\perp  \right) \} \, .   
\end{equation}
With the choices of $t^\pm  = T_1 \pm  T_3$ and $t^\perp = T_2$ (i.e. $\beta_1= \beta_3=0, \beta_2 = 1$) we construct a set of 8 distinct non-invertible lagrangians given in Table \ref{tab:eightdefects}.
  The fusion of these using formula \eqref{eq:liefusion}  is given in Table \ref{tab:Rdefectfusion}.
\renewcommand*{\arraystretch}{1.50}
\begin{table}[b]
    \centering
   \footnotesize
   \begin{align*}
  \begin{array}{cccc}
\circled{green}{1} : & \{ u t^+ + w t^\perp , v t^{-}+ w t^\perp \}  & \circled{green}{2} : & \{ u t^+ + w t^\perp , v t^{+}+ w t^\perp \}  \\  
\circled{blue}{3} : & \{ u t^- + w t^\perp , v t^{-}+ w t^\perp \}     & \circled{blue}{4} : & \{ u t^- + w t^\perp , v t^{+}+ w t^\perp \} \\
\circled{yellow}{5} : & \{ u t^+ + w t^\perp , v t^{-}- w t^\perp \}  & \circled{yellow}{6} : & \{ u t^+ + w t^\perp , v t^{+}- w t^\perp \} \\
\circled{red}{7} : &  \{ u t^- + w t^\perp , v t^{-}- w t^\perp \}    & \circled{red}{8} : & \{ u t^- + w t^\perp , v t^{+}- w t^\perp \} 
\end{array}
\end{align*}
\caption{Some non-invertible lagrangians (see text). }  
    \label{tab:eightdefects}
\end{table}

\begin{table}[]
    \centering
    \renewcommand*{\arraystretch}{1}
\begin{tabular}{|c|c c | c c | c c | c c |}
    \hline
    & \circled{green}{1} & \circled{green}{2} & \circled{blue}{3} & \circled{blue}{4} & \circled{yellow}{5} & \circled{yellow}{6} & \circled{red}{7} & \circled{red}{8} \\ 
    \hline
    \circled{green}{1} &\cellcolor{green!50}{1} & \cellcolor{green!50}{2} & \cellcolor{green!50}{1} & \cellcolor{green!50}{2} & \cellcolor{yellow!50}{5} & \cellcolor{yellow!50}{6} & \cellcolor{yellow!50}{5} &\cellcolor{yellow!50}{6}  \\
    \circled{green}{2} & \cellcolor{green!50}{1} & \cellcolor{green!50}{2} & \cellcolor{green!50}{1} & \cellcolor{green!50}{2} & \cellcolor{yellow!50}{5} & \cellcolor{yellow!50}{6}  & \cellcolor{yellow!50}{5} &\cellcolor{yellow!50}{6}  \\
    \hline
    \circled{blue}{3} & \cellcolor{blue!50}{3} & \cellcolor{blue!50}{4} & \cellcolor{blue!50}{3} & \cellcolor{blue!50}{4}  & \cellcolor{red!50}{7}  & \cellcolor{red!50}{8} &\cellcolor{red!50}{7}  & \cellcolor{red!50}{8} \\
    \circled{blue}{4} & \cellcolor{blue!50}{3}& \cellcolor{blue!50}{4}  & \cellcolor{blue!50}{3} & \cellcolor{blue!50}{4}  &\cellcolor{red!50}{7} & \cellcolor{red!50}{8} & \cellcolor{red!50}{7} & \cellcolor{red!50}{8}  \\
    \hline
    \circled{yellow}{5} & \cellcolor{yellow!50}{5} & \cellcolor{yellow!50}{6} & \cellcolor{yellow!50}{5}  & \cellcolor{yellow!50}{6} & \cellcolor{green!50}{1} &\cellcolor{green!50}{2}  & \cellcolor{green!50}{1} & \cellcolor{green!50}{2} \\
    \circled{yellow}{6} & \cellcolor{yellow!50}{5} &\cellcolor{yellow!50}{6}  & \cellcolor{yellow!50}{5}  & \cellcolor{yellow!50}{6} & \cellcolor{green!50}{1} &\cellcolor{green!50}{2}  & \cellcolor{green!50}{1} &\cellcolor{green!50}{2} \\
    \hline
    \circled{red}{7} & \cellcolor{red!50}{7} &\cellcolor{red!50}{8}  & \cellcolor{red!50}{7} &\cellcolor{red!50}{8} & \cellcolor{blue!50}{3} & \cellcolor{blue!50}{4} & \cellcolor{blue!50}{3} & \cellcolor{blue!50}{4} \\
    \circled{red}{8} & \cellcolor{red!50}{7} & \cellcolor{red!50}{8}  & \cellcolor{red!50}{7} &\cellcolor{red!50}{8}  & \cellcolor{blue!50}{3}  & \cellcolor{blue!50}{4}  & \cellcolor{blue!50}{3} & \cellcolor{blue!50}{4} \\
    \hline
\end{tabular}
\caption{Fusion of $\cR$-defects. This table describes a direct product $\mathbb Z_2\times \mathbb B_2\times \mathbb B_2^\text{op}$ where $\mathbb B_2$ and $\mathbb B_2^\text{op}$ denote the two 2-element band semigroups  (the subset of greens alone defines $\mathbb{B}_2^\text{op}$, whilst blockwise green and blue yield $\mathbb{B}_2$).}
\label{tab:Rdefectfusion}
\end{table}
\renewcommand*{\arraystretch}{1}

Our examples illustrate  that fusion endows the manifold of lagrangians (which is diffeomorphic to $O(d,\mathbb R)$) for a split-signature symmetric form on $\mathbb R^{2d}$   with a \emph{semi-group structure} depending on the signature of $\kappa$. If $\kappa$ is positive-definite, this semigroup is in fact a group, the group $O(d;\mathbb R)$. Non-invertibility occurs when $\kappa$ admits null (lightlike, or isotropic) vectors, i.e.~for \emph{indefinite} signature and has a direct physical interpretation: e.g.~the lagrangian $\frak a^+$ \eqref{eq:2dlagrangians} relevant for defects between $U(1)^2$ CS theories imposes the boundary conditions
\begin{equation}
    A_{N1}=A_{N2}\,,\qquad A_{S1}=-A_{S2} \, , 
\end{equation}
where the gauge fields $A_N$ and $A_S$ do not glue across $D$ so that the topological boundary condition $\frak a^+$ in this folded CS theory unfolds to a pair of topological boundary conditions, one for each side of the defect.

The algebraic structure of the semi-group of lagrangians under fusion is thus controlled by the signature of $\kappa$, i.e.~$s$ pluses, $t\leq s$ minuses. The semi-group is a union of  sub-semi-groups $\mathrm S_{0},\mathrm S_{1},\dots  \mathrm S_{s-t}$ whose fusions obey $\mathrm{S_n}\circ\mathrm{S_m}\subseteq \mathrm{S_{\max(m,n)}}$, of which $S_0=O(\kappa)$ is a group of invertibles and $S_{\mathrm n>1}$ are semi-groups consisting of the non-invertibles.

\section{Quantization conditions}

For defects between abelian CS theories with compact gauge groups $G_N= G_S=U(1)^d$,   one demands that a lagrangian $\frak h$ exponentiates into a compact \footnote{Non-compact subgroups of $U(1)^{2d}$ are not closed topologically; e.g.~a line of irrational slope $\mathbb R\to U(1)^2$ which is dense in the torus. This complicates the reduction of structure group from $U(1)^{2d}$ on the defect.}  
 subgroup  $H =U(1)^{d}$ of $G_N\times G_S=U(1)^{2d}$.  For invertible lagrangians, which take the form  $\{(Mx,x)\mid x\in\frak g_N\}$ for $M\in O(\kappa)$ classically, one would expect that this quantization condition would select lagrangians in $O(\kappa,\mathbb Z)$.

 It does not. To illustrate, consider $d=1$, $\kappa=1$ (we anticipate the following extends to higher dimensions). The allowed lagrangians are
      $\mathfrak h_{m,\pm}=\{m(x,\pm x)\mid x\in\mathbb R \}$ for $m\in\mathbb Z^\times$, where the generators $(1,0)$ and $(0,1)$ are basis vectors for the lattices defining $G_N= G_S=\mathbb R/\mathbb Z$. $O(\kappa)=O(1)$ does not generate the allowed lagrangians.
 
A subset of $O(1,1;\mathbb Q)$ does instead. In the $(1,\pm 1)$ basis, $SO(1,1;\mathbb Q)$ is the $2\times 2$ matrix $\operatorname{diag}(r,r^{-1})$ for $r\in\mathbb Q$ which acts as $(1,1)\mapsto r(1, 1)$ which along with the map $(1,1)\mapsto (1,-1)$ generates the collection of $\mathfrak h_{m,\pm}$.
 However, in general the full action of $O(1,1;\mathbb Q)$  \emph{does not}  respect the quantization.     The fusion of eq.~\eqref{eq:liefusion} \emph{does not} preserve the quantization condition in general.

\section{\label{sec:Abelian} $\mathcal{R}$-defects}

Let us now turn to the tools we will need to exhibit topological defects in non-Abelian CS.  The additional requirement that the lagrangians be subalgebras (compatibility with Gauss' law) is restrictive; whilst the diagonal
\begin{equation}\label{eq:diag_Lagr} 
    \fg_{\Delta} = \{ (x,x) \in \fd \mid   x \in \frak{g}\} \ ,
\end{equation}
is a subalgebra of $\fd$, the anti-diagonal (though isotropic) is not.  To rectify this, we equip $\fg$  with  an endomorphism $\cR$, skew symmetric with respect to $\kappa$,   for which  the $c^2=1$ modified classical Yang-Baxter equation (mCYBE)
\begin{equation} \label{eq:MCYBE}
    [\cR x \, , \cR y] - \cR \big( [\cR x \, , y] + [x\, , \cR y] \big) = - c^2 [x \, , y]  \, , 
\end{equation}
holds for $x,y\in \fg$. We then form a lagrangian subalgebra 
\begin{equation}
    \fg_{\cR} = \{ \big( (\cR + 1) X, (\cR - 1) X \big) \in \fd \} \ .
\end{equation}

  Since its intersection with $\fg_{\Delta}$ is trivial, we have the vector space decomposition $  
    \fd = \fg_{\Delta} \oplus \fg_{\cR} $.
    To construct appropriate defect actions we employ the projectors 
\begin{equation}
    P_{\Delta} =  1 - P_{\mathcal{R}} =  \frac{1}{2} \begin{pmatrix}
        1 - \mathcal{R} & 1 + \mathcal{R} \\
        1 - \mathcal{R} & 1 + \mathcal{R} 
    \end{pmatrix} \, . 
\end{equation}
We let $\bar{\cR} = - \cR $, which also solves the mCYBE, and denote its corresponding lagrangian $\fg_{\bar{\cR}}$.

  For compact semisimple $\fg$ there are no solutions to the the $c^2=1$ mCYBE, whilst the \textit{Drinfeld-Jimbo} (DJ) $\cR$-matrix  provides a canonical solution for the split-signature real form of any complex Lie algebra $\fg^{\bC}$.   
In a  Cartan-Weyl  basis for  $\fg^{\bC}$ this acts as 
\begin{equation} \label{eq:DJ_Rmatrix}
    \cR (H_{i}) = 0 \ , \quad
    \cR (E_{\alpha}) = E_{\alpha} \ , \quad
    \cR (E_{-\alpha}) = -E_{-\alpha} \ .
\end{equation}
$\mathcal R$ descends to the split real form ($\fg$), and enjoys $\cR^3= \cR$.

The lagrangian $  \fg_{\cR}$  has non-trivial fusion properties which we can relate to the  defects  given in Table \ref{tab:eightdefects} with the following identifications:\renewcommand*{\arraystretch}{1.50}
\begin{align*} 
&\circled{green}{1} \> : \quad  \cR \circ \cR \qquad \qquad \;\;\,\;
\circled{green}{2} \> : \quad \cR \circ \bar{\cR}   \qquad \qquad  \qquad \\ &  
\circled{blue}{3} \> : \quad \bar{\cR} \circ \cR \qquad \qquad \;\;\,\,\,
\circled{blue}{4} \> : \quad  \bar{\cR}\circ \bar{\cR} \qquad \qquad \;\;\,\;\, \\ &
\circled{yellow}{5} \> : \quad  \cR  \qquad \qquad \;\;\,\,\quad\;\;\;\,
\circled{yellow}{6} \> : \quad  \cR\circ \cR \circ \bar{\cR} \qquad \qquad \;\;\,\, \\ &
\circled{red}{7} \> : \quad  \bar{\cR}\circ \cR \circ \cR  \qquad  \;\;\;\;\,
\circled{red}{8} \> : \quad \bar{\cR} \qquad 
\end{align*}
\renewcommand*{\arraystretch}{1}

It is useful to introduce two involutive automorphisms on $\fd $.  The first, $\cJ$, acts by swapping `north' and `south' i.e.~$\cJ: (x,y) \mapsto (y, x)$.  The second,  $\cW  (x,y) \mapsto (x,  \textsf{w}(y) )$, uses the Weyl  automorphism, $\textsf{w}: H_i \mapsto - H_i, \, E_{\alpha } \mapsto -E_{-\alpha} 
$.  These act as
\[
\begin{tikzcd}
    {\mathcal R }\ar[r,leftrightarrow,"\mathcal W"] \ar[d,"\mathcal J",leftrightarrow]
    &{\mathcal R\circ \bar{\mathcal R}}\hspace{-.1cm} 
    \ar[loop right, distance=1.3em, out=12, in=-12, "\mathcal J",leftrightarrow]\\
    {\bar{\mathcal R}}\ar[r,leftrightarrow,"\mathcal W"] 
    &{\bar{\mathcal R}\circ\mathcal R}\hspace{-.1cm}
    \ar[loop right, distance=1.3em, out=12, in=-12, "\mathcal J",leftrightarrow]
\end{tikzcd}
\quad
\begin{tikzcd}
    {\mathcal R\circ \mathcal R }\ar[r,leftrightarrow,"\mathcal W"] \ar[d,"\mathcal J",leftrightarrow]
    &{\mathcal R\circ \mathcal R \circ \bar{\mathcal R}}\hspace{-.1cm} 
    \ar[loop right, distance=1.3em, out=8, in=-8, "\mathcal J",leftrightarrow]\\
    {\bar{\mathcal R}\circ\bar{\mathcal R}}\ar[r,leftrightarrow,"\mathcal W"] 
    &{\bar{\mathcal R}\circ\mathcal R\circ\mathcal R}\hspace{-.1cm}
    \ar[loop right, distance=1.3em, out=8, in=-8, "\mathcal J",leftrightarrow]
\end{tikzcd}
\]
 where we use $ \mathcal{R} = \fg_{\mathcal{R}} $ etc. The relations $\cR\circ\cR\circ \cR=\cR$ and $\bar{\cR}\circ \bar{\cR}\circ \bar{\cR}=\bar \cR$  hold, c.f.~$\cR^3=\cR$ and ensure closure of the defects.

\section{Applications to 3D gravity}

Three-dimensional gravity admits a formulation in terms of CS theory \cite{Achucarro:1986uwr,Witten:1988hc}. The Einstein-Hilbert action for negative cosmological constant is equivalent to
\begin{align}
      S_\mathrm{CS}[A]-S_\mathrm{CS}[\bar A]+    \int_{\pd M}  \mathrm{tr}(A\wedge \bar A) \,.
\end{align}
in which the $\frak{sl}_2$ connections are related to the dreibein and dualized spin connection  according to $A^a = \omega^a + e^a$ and $\bar{A} = \omega^a - e^a$.   The boundary term here reproduces the Gibbons-Hawking-York term.     To ensure  a well-defined variational principle that keeps fixed the metric on the boundary a further contribution is required.  In Fefferman-Graham (FG) gauge the combined boundary terms are \cite{Apolo:2015zxh,Llabres:2019jtx,Ebert:2022cle}
\begin{equation}\label{eq:Sbdy}
    S_{\mathrm{bdy}}  =   \int_{\pd M} \mathrm{tr}(A\wedge \bar A) -   \mathrm{tr}((A - \bar A) \wedge  (A - \bar A) L_0 )   \, , 
\end{equation} 
 with $\frak{sl}_2$  generators       obeying $[L_m, L_n] = (m-n) L_{m+n}$.

This can be related to our discussion above by noting that $\cR(\bullet) =- [L_0 , \bullet]$ provides a DJ $\cR$-matrix.   Indeed if we take $\mathbb{A} = (A,\bar{A})$ we have that 
\begin{align*}
  S_\mathrm{CS}[A]-S_\mathrm{CS}[\bar A] +    S_{\mathrm{bdy}} [A,\bar{A}] =   S_\mathrm{CS}[\mathbb{A}] +  \int \biprodb{\mathbb{A}}{P_R^\perp \mathbb{A}} \, .
\end{align*}  
We can immediately extend this to higher spins by replacing $\frak{sl}_2$ with $\frak{sl}_N$; the case of $N=3$ yielding a precise match to the  generalisation of eq.~\eqref{eq:Sbdy} presented in \cite{Apolo:2015zxh}. 
 
Previously, we described defects in a folded theory. Now, we shift focus to asymptotic boundary conditions in the gravitational theory obtained from $\mathcal R$-matrices. From a Chern-Simons perspective, the topological boundary condition $\mathbb{A} \vert_{\pd M} \in \frak{g}_{\cR}$ translates to  
\begin{equation}\label{eq:Rbdycond}
    (\cR - 1) A = (\cR + 1) \bar{A} \quad \Leftrightarrow \quad [e, L_0] = \omega.
\end{equation} 
These differ from the boundary conditions conventionally applied in gravity since we do not impose conditions on individual components of $A,\bar A$ but instead restrict their Lie-algebraic structure.   One then asks: is this boundary condition consistent with asymptotically AdS behavior, and what are the associated asymptotic symmetries?  
 
We view $AdS_3$ as a solid cylinder   with boundary $\partial M =\mathbb{ R}_{t}\times S_\varphi^1$.    In radial gauge fixed form the boundary dynamics is contained in $a(t,\varphi) \sim  A$ and $\bar{a}(t,\varphi) \sim \bar A $. The $\cR$-boundary condition then fixes 
\begin{equation} 
\label{eq:ourboundaryconditions}
 a   =\ell^{+1}L_{+1}+\ell^{0}L_{0}\,,\qquad \bar a =\bar \ell^{-1}L_{-1}-\ell^{0}L_{0}\,, 
\end{equation}
where $\ell^{0},\ell^{+1}$ and $\bar{\ell}^{-1}$  are one-forms on the boundary.  These differ from the general asymptotically $AdS_3$ boundary conditions \cite{Campoleoni:2010zq} which (in diagonal gauge) take the form 
\begin{equation}\label{eq:Andrea}
     a_{AAdS}   = L_{+1}+ \ell^{0}L_{0}  \,,\qquad \bar a_{AAdS}  = L_{-1} + \bar{\ell}^{0}L_{0} \,. 
\end{equation}

In \cite{Campoleoni:2017xyl} it was shown that the asymptotic symmetry algebra corresponding to eq.~\eqref{eq:Andrea} is   $\mathsf{Vir} \times \mathsf{Vir}$ realized in free-field variables. Setting $\ell^{+1} = \bar{\ell}^{-1}=1  $ in  our boundary conditions \eqref{eq:ourboundaryconditions} gives us a special case of the boundary conditions \eqref{eq:Andrea}. Since in \eqref{eq:ourboundaryconditions} $ \ell^{0}=- \bar{ \ell}^{0}$,  only   a diagonal embedding of $\mathsf{Vir}$ survives as the asymptotic symmetry algebra of the boundary condition \eqref{eq:Rbdycond}.

\section{Discussion}

Our construction provides a novel perspective on topological defects in both abelian and non-abelian CS theories. Our methods might point the way towards topological symmetry theories (SymTFTs) for  continuous, and even non-abelian, symmetries (see recent attempts \cite{Brennan:2024fgj,Antinucci:2024zjp,Bonetti:2024cjk}). We speculate this would require a map from the lagrangian subalgebra $\frak h$ in the Drinfel'd double $\frak g\oplus \frak g$, identified here, to lagrangian objects in a representation category of the quantum group arising from that double.

We note that the non-invertibility in our defect fusions is complementary to that observed in e.g.~\cite{Roumpedakis:2022aik}. They find, in the context of abelian CS theories, that some fusions lead to superpositions of defects. We expect that this is related to the quantization condition we outlined in a previous section. One might speculate that non-invertible defects in that sense arise from elements of $O(d,d;\mathbb Q)$ which would lead to connections $\mathbb{A}_{NS}$ to be such that some integer power of its holonomies are in $U(1)^{d} $; similar ideas are considered in \cite{Kapustin:2010hk}.  
In a string-theoretic context the interplay of $O(d,d;\mathbb Q)$ with quantization of lagrangians was considered in \cite{Bachas:2012bj}.

With regard to gravity applications, our $AdS_3$ boundary conditions led to an asymptotic symmetry algebra which is a \emph{single} copy of the Virasoro algebra. This approach might thus be especially interesting for \emph{chiral higher-spin gravity} \cite{Skvortsov:2018jea,Sharapov:2024euk} for which one expects a similar structure.

\section{acknowledgments}
\begin{acknowledgments}
The authors would like to thank Andrea Campoleoni, Thanassis Chatzistavrakidis, Arnaud Delfante, 
 Monica Guica, Ben Hoare, Nabil Iqbal, Larisa Jonke, Prem Kumar, Carlos Nunez, Matt Roberts, Jan Rosseel, Evgeny Skvortsov  and C\'eline Zwikel for invaluable discussions on the topic of this paper and related ideas. A.S.A.~is funded by the Croatian Science Foundation project “HigSSinGG — Higher Structures and Symmetries in Gauge
and Gravity Theories” (IP-2024-05-7921). D.C.T. is supported by the STFC grant ST/X000648/1, and by the Royal Society through a University Research Fellowship `Generalised Dualities in String Theory and Holography' URF 150185.   S.D. is grateful to the Azrieli foundation for the award of an Azrieli fellowship. The work of S.D. is also partially supported by the Israel Science Foundation (grant No. 1417/21), by Carole and Marcus Weinstein through the BGU Presidential Faculty Recruitment Fund, the ISF Center of Excellence for theoretical high energy physics and by the Origins Excellence Cluster and by the German-Israel-Project (DIP) on Holography and the Swampland, and by the ERC Starting Grant dSHologQI (project number 101117338).
 
\end{acknowledgments}

\bibliography{bibliography}
 
\end{document}